\documentclass[aps,pre,superscriptaddress,twocolumn,showkeys]{revtex4-1}

\usepackage{amssymb,latexsym,amsthm,amsmath}    
\usepackage{amsfonts}             

\usepackage{graphicx}                           
\usepackage{color}                             
\usepackage{ulem}

\begin{document}

\title{Topological Measure Locating the Effective Crossover between Segregation and Integration in a Modular Network}
\author{A. Ajdari Rad}
\affiliation{\'Ecole Polytechnique F\'ed\'eral de Lausanne, Laboratory of Nonlinear Systems, School of Computer and Communication Science, 1015 Lausanne, Switzerland}
\author{I. Sendi\~na-Nadal}
\affiliation{Complex Systems Group, Universidad Rey Juan Carlos, 28933 M\'ostoles, Madrid, Spain}
\affiliation{Center for Biomedical
  Technology, Universidad Polit\'ecnica de Madrid, Campus de Montegancedo, 28223 Pozuelo de Alarc\'on, Madrid, Spain  }
\author{D. Papo}
\affiliation{Center for Biomedical
  Technology, Universidad Polit\'ecnica de Madrid, Campus de Montegancedo, 28223 Pozuelo de Alarc\'on, Madrid, Spain  }
\author{M. Zanin}
\affiliation{Center for Biomedical
  Technology, Universidad Polit\'ecnica de Madrid, Campus de Montegancedo, 28223 Pozuelo de Alarc\'on, Madrid, Spain  }
\author{J. M. Buld\'u}
\affiliation{Complex Systems Group, Universidad Rey Juan Carlos, 28933 M\'ostoles, Madrid, Spain}
\affiliation{Center for Biomedical
  Technology, Universidad Polit\'ecnica de Madrid, Campus de Montegancedo, 28223 Pozuelo de Alarc\'on, Madrid, Spain  }
\author{F. del Pozo}
\affiliation{Center for Biomedical
  Technology, Universidad Polit\'ecnica de Madrid, Campus de Montegancedo, 28223 Pozuelo de Alarc\'on, Madrid, Spain  }
\author{S. Boccaletti}
\affiliation{Center for Biomedical
  Technology, Universidad Polit\'ecnica de Madrid, Campus de Montegancedo, 28223 Pozuelo de Alarc\'on, Madrid, Spain  }

\date{\today}


\begin{abstract}
{\color{black}We introduce an easily computable topological measure which
locates the effective crossover between segregation and integration in
a modular network.}
Segregation corresponds to the degree of network modularity, while
integration is expressed in terms of the algebraic connectivity of an
associated hyper-graph. 
{\color{black}The rigorous treatment of the simplified case of cliques
  of equal size that are gradually rewired until they become
  completely merged, allows us to show that this topological crossover can be made to coincide with a
dynamical crossover from cluster to global synchronization of a
system of coupled phase oscillators.
The
dynamical crossover is signaled by a peak in the product of the
measures of intra-cluster and global synchronization, which we propose
as a dynamical measure of complexity. This quantity is much easier to
compute than the entropy (of the average frequencies of the
oscillators), and displays a behavior which closely mimics that of
the dynamical complexity index based on the latter. 
The proposed toplogical measure simultaneously
provides information on the dynamical behavior, sheds
light on the interplay between modularity vs total integration and
shows how this affects the capability of the network to perform both local and
distributed dynamical tasks.}

PACS: 89.75.Fb, 05.45.Xt, 89.70.Eg
\end{abstract}


\maketitle

Many physical and biological systems (such as electronic devices, communications networks, and the human brain) face similar constraints as they interact with complex environments, and organize their structure and function along similar principles of resource allocation \cite{Laughlin2003}. On the one hand, the need for fast and reliable responses to changes in the environment naturally favors the emergence of segregated modules of specialized
computation (e.g. sensory systems in the brain). On the other hand, interactions among modules become essential when an information processing whose complexity exceeds the capacity of the single modules is required. For instance, perceptual systems in the brain need to bind information from different brain areas to produce a single coherent percept \cite{Varela2001}. Therefore, segregation into specialized modules and integration into global coherent activity present an inherent trade-off, and an appropriate balance between these two tendencies has been shown to be necessary for efficient functioning, particularly in neural systems \cite{Tononi1994}. In fact, an exceedingly segregated or integrated functioning of the brain has been associated with various pathological conditions, e.g. autism or schizophrenia \cite{Fletcher1997,Tononi1998,Just}, and epilepsy \cite{Stam2005} respectively.

One straightforward way to study such a balance in complex systems is to represent them as dynamical networks, endowing them with well-studied topological and dynamical properties (see \cite{Boccaletti2006,Arenas2008} for a review). For instance, Zhao {\it et al.} \cite{Zhao2010} characterized systems of coupled phase oscillators in terms of a complexity index based on the entropy of the distribution of pairwise synchronization. Heterogeneous and modular networks were shown to be characterized by high complexity, for intermediate levels of modularity, in a regime marked by the formation of dynamical clusters and the coordination between them.

In this Letter, we provide an easily computable topological measure for quantifying the balance between segregation and integration in a network. We propose that segregation can be understood in terms of a community structure (i.e. clusters of vertices densely connected to each other while less connected to vertices outside the community \cite{Girvan2002}), while integration can conveniently be expressed in terms of algebraic connectivity \cite{algebra} of the hyper-graph associated to the network.
After introducing our measure for a generic modular graph, we focus on
the simplified case in which the network communities are of equal
size, and have the same number of inner and outer connections, and
show analytically that there is a structure that maximizes the product
of segregation and coordination measures. We then demonstrate that the
dynamics emerging from such a specific configuration is associated
with  {\color{black}the coincidence of the two thresholds for cluster and complete
synchronization in a network of interacting phase oscillators}.

We start by considering a generic undirected, un-weighted graph
$\cal{G}$ composed of $N$ nodes and $L$ links, partitioned into $C$
communities, and characterized by an associated $N \times N$ adjacency
matrix $A$. Moreover, for each community $r$ ($r=1,...,C$), we
consider the number of links connecting pairs of members of that
community ($\ell_{in}^r$), and the number of inter-community links
($\ell^r_{out}$), i.e. the number of links connecting a member of that
community with a member of another community. Based on the mentioned notation, we have $L= \left( \sum_{r=1}^{C} \ell^r_{in} \right) + \frac{1}{2} \left( \sum_{r=1}^{C} \ell^r_{out} \right)$.
Using the previous definitions, the standard modularity measure defined in \cite{Newman2004}, can be written as $Q=\sum^C_{r=1}\{\ell_{in}^r/L-[(2\ell_{in}^r+\ell_{out}^r)/2L]^2\}$.

Let us now define the hyper-graph $\cal{G}^*$ associated to $\cal{G}$,
as the weighted directed C-clique in which each node corresponds to a
community of $\cal{G}$, and the connection incident to node $r$ from
node $s$ is weighted by $\frac {\ell^{rs}_{out}}{\ell^r_{in}}$, being
$\ell^{rs}_{out}$ the number of links of $\cal{G}$ that connects
members of the community $r$ with members of the community $s$, 
and $\ell^r_{in}$ the number of inner links in the source community.
The corresponding $C\times C$ Laplacian matrix ${\cal L}^*= \{ {\cal
  L}_{rs}^* \}$ is asymmetric, but can be written as the product
${\cal L}^* = {\cal B} {\cal C}$, where $\cal C$ is a symmetric zero row-sum matrix with off-diagonal elements ${\cal C}_{rs}=-\ell^{rs}_{out}$ and diagonal ones ${\cal C}_{rr} =\ell^r_{out}=\sum_{s\ne
  r}\ell^{rs}_{out}$, and ${\cal B}=\mbox{diag}
\{1/\ell^1_{in},\cdots,1/\ell^C_{in}\}$:

\begin{equation*}
{\cal L^*}=
\left(
\begin{array}{cccc}
\frac{1}{\ell^1_{in}} & 0 & \cdots & 0\\
0 & \frac{1}{\ell^2_{in}} & \cdots & 0\\
\vdots & \vdots & \ddots & \vdots \\
0 & 0 & \cdots & \frac{1}{\ell^C_{in}}\\
\end{array}
\right)
\left(
\begin{array}{cccc}
\ell^1_{out} & -\ell^{12}_{out} & \cdots & -\ell^{1C}_{out}\\
-\ell^{21}_{out} &\ell^{2}_{out} & \cdots &- \ell^{2C}_{out}\\
\vdots & \vdots & \ddots & \vdots \\
-\ell^{C1}_{out} &-\ell^{C2}_{out} & \cdots & \ell^{C}_{out}\\
\end{array}
\right)
\label{matrix}
\end{equation*}

The spectrum of $\cal L^*$ is  real with non-negative values and because ${\cal L}^*$ is zero row-sum, the smallest eigenvalue $\lambda_1^*$ is zero, while $\lambda^*_2>0$.
The measure that we propose for the balance between integration and segregation is defined as follows:

\begin{equation*}
\xi=Q\lambda_2^*.
\label{measure}
\end{equation*}

Indeed, while $Q$ is an inherent evaluation of the segregation factor of a graph, $\lambda^*_2>0$ quantifies the connectiveness of the hyper-graph, and therefore measures the extent to which different communities are bounded and interact. It should be noticed that both $Q$ and $\lambda_2^*$ are properly normalized in such a way that, even if the network links were associated to cohesive forces, the two quantities would be a-dimensional.
The maximum of $\xi$ corresponds to a topology in which integration
and segregation have the same weight. 
\begin{figure}
  \centering\includegraphics[width=0.49\textwidth]{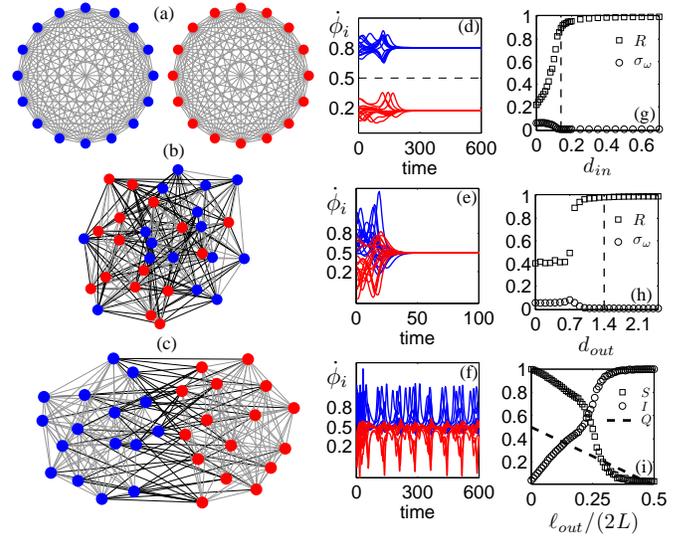}
\caption[]{(Color online) (a)-(c) Graph representations of (a) a network
  consisting of two cliques of size $N_C=16$,  and (b-c) the resulting
  networks after rewiring. (b) $j^{max}=60$
  ($\ell_{out}/2L=0.5$ and $Q=0$), and (c) $j=30$
($\ell_{out}/2L=0.25$ and $Q=0.25$). Intra-community (inter-community) links are light
gray (black colored).
Blue (red) nodes are phase oscillators whose natural frequency is randomly distributed around $0.8$ (0.2)
within a range $\pm 0.1$.
(d)-(f) Time evolution of the instantaneous frequencies of the $N=32$
oscillators after solving
Eq.~(\ref{system}) with the networks of the left panels and $d_{in}=d_{in}^*=0.14$.
(d) $d_{out}=0$, (e)
$d_{out}=d_{out}^*=1.10$, and (f) $d_{out}=1.40$.
Dashed line in (d) marks the mean frequency between the two clusters.
(g)-(h) $R$ (squares) and $\sigma_{\omega}$ (circles) (see text for definition)  for:
(g) the two clique network of (a) as a function of $d_{in}$, (h) the 0-modularity network of (b) as a function of
$d_{out}$ and for $d_{in}=0.14$. The vertical dashed line
in (g) marks the critical coupling for cluster synchronization ($d_{in}^*=0.14$),
while in (h) it corresponds to $d_{out}=1.4$,  well above the onset of global
synchronization ($d_{out}^*=1.10$). Panel (i) reports the $S$ and $I$ indexes (see text for
definitions) vs. the ratio $\ell_{out}/(2L)$, after solution of Eq.~(\ref{system}) with $d_{in}=0.14$ and
$d_{out}=1.40$. Oblique dashed line is the corresponding modularity index $Q$.}
\label{fig1}
\end{figure}

{\color{black} Let us then consider the case in which the $C$ communities are cliques
having  equal size $N_c=N/C$,}  and the number of intra-community links,
$\ell^r_{in}$, as well as the number of inter-community links,
$\ell^r_{out}$, are the same for all communities. We then have
$\ell^r_{in}=\ell_{in}/C$, and $\ell^r_{out}=\ell_{out}/C$, such 
that $L=C(\ell^r_{in} +\frac{1}{2}\ell^r_{out})= \ell_{in} + \frac{1}{2}\ell_{out}$. Under these assumptions, the modularity can be reduced to the following expression,

\begin{equation}   \label{eq:modularity} Q = C\left[
  \frac{1}{L}\frac{\ell_{in}}{C}-\left(\frac{2L}{C}\frac{1}{2L}\right)^2\right]=1-\frac{1}{2L}\ell_{out}-\frac{1}{C}.
\end{equation}

On the other hand,  we have that ${\cal B }=\frac{C}{\ell_{in}}{\cal I}$ and, from matrix identity,

\begin{equation}
  \label{eq:lambda2}
\lambda_2^*= \frac{\ell_{out}}{\ell_{in}}.
\end{equation}

{\color{black}Let us now consider a particular protocol by
means of which $l_{out}$ is varied from 0 to $2L(C-1)/C$, the value at
which the modularity $Q$ is
zero. We start from a fully segregated configuration}
 (in which  $\ell_{out}=0$ and $\ell_{in}=L$), and operate successive rewiring processes, in each of which an intra-community link from each community is deleted, and $C$ inter-community links are formed by connecting those pairs of nodes (each one in different communities) having lost their intra-link. In this way, at the $j$-th rewiring, we have $\ell_{in}=L - Cj$, and $\ell_{out}=C 2 j$. The
maximum number of rewiring steps until modularity fades out is, therefore, $N_c(N_c-1)(C-1)/(2C)$.
{\color{black}It follows from combining Eqs.~(\ref{eq:modularity}) and
  ~(\ref{eq:lambda2}) that $ \xi =  Q  \lambda_2^* =
  \left(1-\frac{1}{C}-\frac{1}{2L}\ell_{out}\right) \frac{\ell_{out}}{L-\frac{1}{2}\ell_{out}}$.
Accordingly, the partial derivative of $\xi$ w.r.t. $\ell_{out}$ is $\frac{\partial \xi}{{\partial \ell}_{out}}= 1-\frac{1}{C}
-\frac{\ell_{out}}{L} + \frac{1}{4}\left(\frac{\ell_{out}}{L}\right)^2$,
which vanishes at 
\begin{equation}
  \label{eq:optimallout}
  \ell_{out}^{max}=2L(1-\frac{1}{\sqrt{C}}),
\end{equation}
\noindent
i.e., there exists a value of inter-community links at which $\xi$
reaches its maximum value (the second derivative of $\xi(\ell_{out})$ is
indeed negative)}.

\begin{figure}
  \centering
 \includegraphics[width=0.44\textwidth]{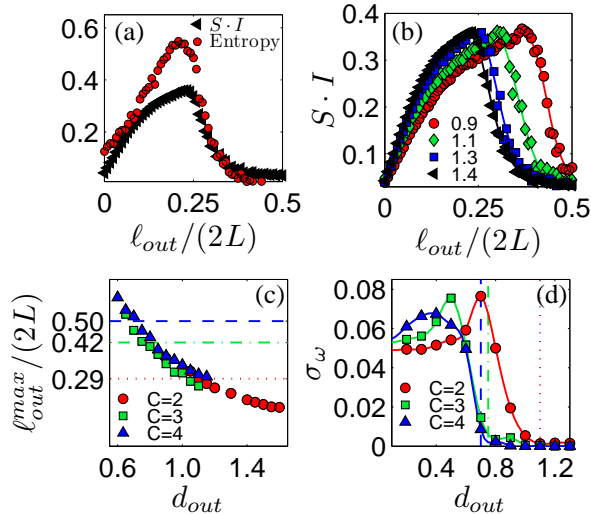}
\caption[]{(Color online) (a) Complexity index $E$ (circles) and $S\cdot I$ index (triangles) (see text for
definitions) corresponding to the data reported in Fig.~\ref{fig1}(i). (b) $S\cdot I$ curves as a function of the mixing
ratio $\ell_{out}/(2L)$ for $C=2$ and for increasing values of
$d_{out}$ (in the legend).
(c) Maxima of the $S\cdot I$ curves in (b) vs. $d_{out}$ for $C=2$, $3$,
and $4$. Horizontal lines mark the maxima provided by
Eq.~(\ref{eq:optimallout}): $C=2$ (red dotted), $C=3$ (green dash dotted), and $C=4$ (blue
dashed). Each line intersects the corresponding $S\cdot I$ maxima curve
at a particular $d_{out}$, that coincides with the threshold of complete synchronization,
as shown by the corresponding vertical lines in (d).}\label{fig2}
\end{figure}

We now show that the maximization of $\xi$ has a dynamical counterpart for the analytically treated example.
To this purpose, we consider a modular network of $N$ nodes
initially arranged in two cliques $c_1$ and $c_2$ of equal size $N_c=\frac{N}{2}$ (Fig.~\ref{fig1}(a)), with $\ell^{c_1}_{out}=\ell^{c_2}_{out}=0$ and
$\ell^{c_1}_{in}=\ell^{c_2}_{in}=N_c(N_c-1)/2$ such that the total number of links is $L=N_c(N_c-1)$.
The units of the network are taken to be phase oscillators evolving according to the Kuramoto model \cite{Kuramoto1984}:
\begin{equation}
  \label{system}
  \dot\phi_i=
  \omega_{i}+\frac{1}{N_c-1}\sum_{j=1}^N d_{ij}a_{ij}\sin(\phi_j-\phi_i)
\end{equation}
\noindent
where $\dot\phi_i$ is the angular velocity of the $i$-th oscillator, $\omega_i$ its natural frequency,
$d_{ij}$ is the coupling
strength between oscillators $i$ and $j$ (in the following $d_{ij}=d_{in}$ if the nodes $i$ and
$j$ are members of the same community, and $d_{ij}=d_{out}$ otherwise), and $a_{ij}=1$
if $i$ and $j$ are connected and
$0$ otherwise. Oscillators
$i=1,\dotsc,N_c$ ($i=N_c+1,\dotsc,N$), initially assigned to clique
$c_1$ ($c_2$) and colored in blue (red) in Fig.~\ref{fig1}(a-c),
have $\omega_i$ uniformly distributed around
$\omega_{c_1}=0.8$ ($\omega_{c_2}=0.2$) within a range of $\pm 0.1$.

To explore the extent of validity of
Eq.~(\ref{eq:optimallout}), we start from such a completely segregated configuration, and gradually increase
 the ratio $\ell_{out}/2L$ by {\it i)} randomly selecting a link in each
module, {\it ii)} deleting it, and {\it iii)} forming two new inter-community links by
pairwise connecting the ends of the deleted links. The rewiring process is then
repeated up to the point in which the modularity
index $Q$ cancels, and both modules become statistically indistinguishable
(Fig.~\ref{fig1}(b)). According to
Eq.~(\ref{eq:optimallout}), the balance for a $C=2$ module
network is found at the ratio $\ell_{out}/2L=1-\sqrt{1/2}$.

Furthermore, the tuning of the coupling strengths $d_{in}$ and  $d_{out}$ allows one to control the degree
of intra-community (cluster) and inter-community (complete) synchronization in each given
configuration. Namely, we compute the order parameter
$R=\langle \parallel 1/N \sum_{j=1}^N e^{i\phi_j}\parallel \rangle_T$
(squares in Figs.~\ref{fig1}(g-h)) and the frequency standard deviation $\sigma_{\omega}$
(circles in Figs.~\ref{fig1}(g-h)), averaged over a proper time window $T$, to account for the phase and frequency synchronization (in our simulations $T=200$ t.u.).
In Fig.~\ref{fig1}(g), $d_{in}$ is
increased up to $d_{in}^*=0.14$, that constitutes
the threshold for cluster synchronization in the case of the
network structure of Fig.~\ref{fig1}(a). Fig.~\ref{fig1}(d) shows that
the instantaneous frequencies $\dot\phi_i$ of all
oscillators are locked to their respective cluster frequencies,
$0.2$ (in red) and $0.8$ (in blue). On the other hand, when the modularity index vanishes
(as in the network of Fig.~\ref{fig1}(b)), we set $d_{in}$ to
$d_{in}^*$ and vary $d_{out}$ to find the threshold for global
synchronization at $d_{out}^*=1.10$ (Fig.~\ref{fig1}(h)).
As shown in Fig.~\ref{fig1}(e), for $d_{out}=1.40$ (sufficiently above the
transition), the network is performing a collective oscillation
at the mean frequency of the two clusters ($0.5$).

This way, functional segregated and integrated states are guaranteed by
choosing $d_{in}$ and $d_{out}$ above the threshold couplings
for cluster synchronization and global synchronization. Depending on
the fraction of inter-community links, the competition between
the dynamics of the individual clusters and the whole network will
give rise to a certain degree of functional segregation and
integration. Figure ~\ref{fig1}(f) shows the instantaneous
frequencies of the oscillators coupled according to the scheme of
Fig.~\ref{fig1}(c) for $\ell_{out}/2L=0.25$. One can easily see that the
behavior is far from being totally integrated or segregated but,
instead, the instantaneous frequencies of the oscillators undergo high amplitude oscillations
around their natural values.

To quantify the degree of dynamical segregation $S$ and dynamical
integration $I$, we calculate the
ensemble averages $S=\langle S_i \rangle$ and $I=\langle I_i\rangle$.
$S_i$ and $I_i$  are the dynamical segregation and integration measures for the
individual $i$ oscillators, and are defined as
$S_i=\langle\lvert \dot\phi_i-\frac{\omega_{c_1}+\omega_{c_2}}{2}\lvert\rangle_{T}$
(the time-averaged absolute distance between the oscillator's instantaneous
frequency and the mean frequency of the two clusters), and $I_i=\langle\lvert
\dot\phi_i -\delta(s_i,c_1)\omega_{c_{1}}-\delta(s_i,c_2)\omega_{c_2}\lvert\rangle_{T}
$, where $s_i$ is the community of which the oscillator $i$ is a member,
and $\delta (s_i,s_j)$ the Kronecker delta function.
The dependence of these quantities on the mixing ratio $\ell_{out}/2L$
is then normalized, and reported in Fig.~\ref{fig1}(i) for $d_{in}=d_{in}^*$ and
$d_{out}=1.40>d_{out}^*$, showing a monotonous decreasing behavior of $S$ (squares),
as the modularity $Q$ (dashed line) vanishes, while a monotonous increasing
trend of $I$ (circles), up to saturation when the whole network is
fully synchronized.

Consequently, the dynamical segregation/integration trade-off, measured as the
product of $S$ and $I$, gives information about the existence of a
level of topological mixing for which functional clustering
balances global synchronization, as shown in Fig.\ref{fig2}(a)
(triangles). An alternative way to measure the combination of
dynamical segregation and integration is by means of the complexity
index $E$, introduced in Ref.~\cite{Zhao2010} in the context of
oscillatory networks. Here, $E$ is calculated using
the Shannon entropy of the distribution $P(\omega)$ of the
average frequencies of all oscillators
as $E=(-\sum_{l=1}^m P_l \ln P_l)/ \ln m$, where $m$ is
the number of bins in the histogram of $P(\omega)$. By
definition, $E$ should be close to zero for narrow
distributions, while it should take large values for broad distributions
reflecting the emergence of
complexity. The index $E$ is plotted in Fig.~\ref{fig2}(a)
(full circles) together with the product of the dynamical indexes $S$
and $I$ showing a noticeably similar behavior. It is important to remark that
the calculation of $S \cdot I$ implies much simpler operations, once
the output of system (\ref{system}) is available.

To more closely inspect the relation between topology and
dynamics, we study in Fig.~\ref{fig2}(b) the influence of the coupling $d_{out}$ on the  balance
between segregation and integration. The main observation is that the
 number of inter-community links $\ell_{out}$ that compromises the
balance between the two competing synchronization processes decreases as
the link strength $d_{out}$ increases. We confirm that this
trend also holds for modular networks with $C>2$. Namely, we
have constructed three and four clique networks and performed the same
analysis done for $C=2$, obtaining the fraction $\ell_{out}/2L$
at which the maxima of the $S \cdot I$ curves occur. Results are
reported in Fig.~\ref{fig2}(c) for the three values of $C$ (triangles
for $C=4$, squares for $C=3$, and circles for $C=2$). Therefore,
the functional balance depends on $d_{out}$ and arises for a given
ratio $\ell_{out}/2L$ where the frequency synchronization within the
modules is still effective and, at the same time there is significant
coordination between the modules characterized by the much richer
behavior with the presence of several time scales {\color{black}(as it
  can be observed from Fig.~\ref{fig2}(a) where the entropy of the
  distribution of frequencies exhibits a maximum)}.
However, the balance between modularity and algebraic connectivity is
maximum at one particular value of the mixing ratio as expressed by
Eq.~(\ref{eq:optimallout}) (see Fig.~\ref{fig2}(c)).
Strikingly, the intersection of the analytical $C$
line $\ell_{out}^{max}/2L=1-1/\sqrt{C}$ with the curve of the maxima
location of the $S\cdot I$ curve predicts the value of $d_{out}$ that perfectly coincides with
the coupling threshold for the onset of the global synchronization, as
shown by the vertical lines of Fig.~\ref{fig2}(d). Qualitatively similar scenarios
have been observed for 3D chaotic oscillators, and for 2D excitable units.

{\color{black}In conclusion, we introduced an easily computable topological
  measure from the knowledge of the adjacency matrix of the graph, and
  from a given partition, which locates the effective crossover between segregation
  and integration in a modular network, and shows that its maximum
   has a dynamical counterpart in the tradeoff between the onsets of
   cluster and complete synchronization of networked phase oscillators.}  Our results can therefore enlighten over the performance
of biological systems, that have to organize their structure and
function to simultaneously perform specialized computations at smaller
scales and bind information at larger ones.
 As such, it
can be applied, in principle, to any real world modular network,
namely to evaluate the extent to which a specific configuration has
been optimized for information processing, as well as in connection with genetic,
or simulated annealing  algorithms, for the generation of the optimal
modular structure with a given number of nodes and links.

Work supported by the Ministerio de Educaci\'on y Ciencia of Spain
(FIS2009-07072) and by the Community of Madrid under projects
URJC-CM-2010-CET-5006 and R\&D Program MODELICO-CM (S2009ESP-1691). Authors acknowledge the computational resources and assistance provided by CRESCO, the center of ENEA in
Portici, Italy.


\begin{thebibliography}{14}
\expandafter\ifx\csname natexlab\endcsname\relax\def\natexlab#1{#1}\fi
\expandafter\ifx\csname bibnamefont\endcsname\relax
  \def\bibnamefont#1{#1}\fi
\expandafter\ifx\csname bibfnamefont\endcsname\relax
  \def\bibfnamefont#1{#1}\fi
\expandafter\ifx\csname citenamefont\endcsname\relax
  \def\citenamefont#1{#1}\fi
\expandafter\ifx\csname url\endcsname\relax
  \def\url#1{\texttt{#1}}\fi
\expandafter\ifx\csname urlprefix\endcsname\relax\def\urlprefix{URL }\fi
\providecommand{\bibinfo}[2]{#2}
\providecommand{\eprint}[2][]{\url{#2}}

\bibitem[{\citenamefont{Laughlin and Sejnowski}(2003)}]{Laughlin2003}
\bibinfo{author}{\bibfnamefont{S.}~\bibnamefont{Laughlin}} \bibnamefont{and}
  \bibinfo{author}{\bibfnamefont{T.}~\bibnamefont{Sejnowski}},
  \bibinfo{journal}{Science} \textbf{\bibinfo{volume}{301}},
  \bibinfo{pages}{1870} (\bibinfo{year}{2003}).

\bibitem[{\citenamefont{Varela et~al.}(2001)\citenamefont{Varela, Lachaux,
  Rodriguez, and Martinerie}}]{Varela2001}
\bibinfo{author}{\bibfnamefont{F.}~\bibnamefont{Varela}},
  \bibinfo{author}{\bibfnamefont{J.}~\bibnamefont{Lachaux}},
  \bibinfo{author}{\bibfnamefont{E.}~\bibnamefont{Rodriguez}},
  \bibnamefont{and}
  \bibinfo{author}{\bibfnamefont{J.}~\bibnamefont{Martinerie}},
  \bibinfo{journal}{Nat. Rev. Neurosci.} \textbf{\bibinfo{volume}{2}},
  \bibinfo{pages}{229} (\bibinfo{year}{2001}).

\bibitem[{\citenamefont{Tononi et~al.}(1994)\citenamefont{Tononi, Sporns, and
  Edelman}}]{Tononi1994}
\bibinfo{author}{\bibfnamefont{G.}~\bibnamefont{Tononi}},
  \bibinfo{author}{\bibfnamefont{O.}~\bibnamefont{Sporns}}, \bibnamefont{and}
  \bibinfo{author}{\bibfnamefont{G.~M.} \bibnamefont{Edelman}},
  \bibinfo{journal}{Proc. Natl. Acad. Sci.} \textbf{\bibinfo{volume}{91}},
  \bibinfo{pages}{5033} (\bibinfo{year}{1994}).

\bibitem[{\citenamefont{Fletcher et~al.}(1999)\citenamefont{Fletcher, McKenna,
  Friston, Frith, and Dolan}}]{Fletcher1997}
\bibinfo{author}{\bibfnamefont{P.}~\bibnamefont{Fletcher}},
  \bibinfo{author}{\bibfnamefont{J.}~\bibnamefont{McKenna}},
  \bibinfo{author}{\bibfnamefont{K.}~\bibnamefont{Friston}},
  \bibinfo{author}{\bibfnamefont{C.}~\bibnamefont{Frith}}, \bibnamefont{and}
  \bibinfo{author}{\bibfnamefont{R.}~\bibnamefont{Dolan}},
  \bibinfo{journal}{Neuroimage} \textbf{\bibinfo{volume}{9}},
  \bibinfo{pages}{337} (\bibinfo{year}{1999}).

\bibitem[{\citenamefont{Tononi et~al.}(1998)\citenamefont{Tononi, McIntosh,
  Russell, and Edelman}}]{Tononi1998}
\bibinfo{author}{\bibfnamefont{G.}~\bibnamefont{Tononi}},
  \bibinfo{author}{\bibfnamefont{A.~R.} \bibnamefont{McIntosh}},
  \bibinfo{author}{\bibfnamefont{D.~P.} \bibnamefont{Russell}},
  \bibnamefont{and} \bibinfo{author}{\bibfnamefont{G.~M.}
  \bibnamefont{Edelman}}, \bibinfo{journal}{NeuroImage}
  \textbf{\bibinfo{volume}{7}}, \bibinfo{pages}{133 } (\bibinfo{year}{1998}).

\bibitem[{\citenamefont{Just et~al.}(2004)\citenamefont{Just, Cherkassky, T.A.,
  and Minshew}}]{Just}
\bibinfo{author}{\bibfnamefont{M.}~\bibnamefont{Just}},
  \bibinfo{author}{\bibfnamefont{V.}~\bibnamefont{Cherkassky}},
  \bibinfo{author}{\bibfnamefont{K.}~\bibnamefont{T.A.}}, \bibnamefont{and}
  \bibinfo{author}{\bibfnamefont{N.}~\bibnamefont{Minshew}},
  \bibinfo{journal}{Brain} \textbf{\bibinfo{volume}{127}},
  \bibinfo{pages}{1811} (\bibinfo{year}{2004}).

\bibitem[{\citenamefont{Stam}(2005)}]{Stam2005}
\bibinfo{author}{\bibfnamefont{C.}~\bibnamefont{Stam}},
  \bibinfo{journal}{Clinical neurophysiology} \textbf{\bibinfo{volume}{116}},
  \bibinfo{pages}{2266} (\bibinfo{year}{2005}).

\bibitem[{\citenamefont{Boccaletti et~al.}(2006)\citenamefont{Boccaletti,
  Latora, Moreno, Chavez, and Hwang}}]{Boccaletti2006}
\bibinfo{author}{\bibfnamefont{S.}~\bibnamefont{Boccaletti}},
  \bibinfo{author}{\bibfnamefont{V.}~\bibnamefont{Latora}},
  \bibinfo{author}{\bibfnamefont{Y.}~\bibnamefont{Moreno}},
  \bibinfo{author}{\bibfnamefont{M.}~\bibnamefont{Chavez}}, \bibnamefont{and}
  \bibinfo{author}{\bibfnamefont{D.-U.} \bibnamefont{Hwang}},
  \bibinfo{journal}{Phys. Rep.} \textbf{\bibinfo{volume}{424}},
  \bibinfo{pages}{175 } (\bibinfo{year}{2006}).

\bibitem[{\citenamefont{Arenas et~al.}(2008)\citenamefont{Arenas,
  D{\'i}az-Guilera, Kurths, Moreno, and Zhou}}]{Arenas2008}
\bibinfo{author}{\bibfnamefont{A.}~\bibnamefont{Arenas}},
  \bibinfo{author}{\bibfnamefont{A.}~\bibnamefont{D{\'i}az-Guilera}},
  \bibinfo{author}{\bibfnamefont{J.}~\bibnamefont{Kurths}},
  \bibinfo{author}{\bibfnamefont{Y.}~\bibnamefont{Moreno}}, \bibnamefont{and}
  \bibinfo{author}{\bibfnamefont{C.}~\bibnamefont{Zhou}},
  \bibinfo{journal}{Phys. Rep.} \textbf{\bibinfo{volume}{469}},
  \bibinfo{pages}{93 } (\bibinfo{year}{2008}).

\bibitem[{\citenamefont{Zhao et~al.}(2010)\citenamefont{Zhao, Zhou, Chen, Hu,
  and Wang}}]{Zhao2010}
\bibinfo{author}{\bibfnamefont{M.}~\bibnamefont{Zhao}},
  \bibinfo{author}{\bibfnamefont{C.}~\bibnamefont{Zhou}},
  \bibinfo{author}{\bibfnamefont{Y.}~\bibnamefont{Chen}},
  \bibinfo{author}{\bibfnamefont{B.}~\bibnamefont{Hu}}, \bibnamefont{and}
  \bibinfo{author}{\bibfnamefont{B.-H.} \bibnamefont{Wang}},
  \bibinfo{journal}{Phys. Rev. E} \textbf{\bibinfo{volume}{82}},
  \bibinfo{pages}{046225} (\bibinfo{year}{2010}).

\bibitem[{\citenamefont{Girvan and Newman}(2002)}]{Girvan2002}
\bibinfo{author}{\bibfnamefont{M.}~\bibnamefont{Girvan}} \bibnamefont{and}
  \bibinfo{author}{\bibfnamefont{M.~E.~J.} \bibnamefont{Newman}},
  \bibinfo{journal}{Proc. Natl. Acad. Sci.} \textbf{\bibinfo{volume}{99}},
  \bibinfo{pages}{7821} (\bibinfo{year}{2002}).

\bibitem[{\citenamefont{Chung}(1997)}]{algebra}
\bibinfo{author}{\bibfnamefont{F.}~\bibnamefont{Chung}},
  \emph{\bibinfo{title}{Spectral Graph Theory}}, vol.~\bibinfo{volume}{92} of
  \emph{\bibinfo{series}{CBMS Regional Conference Series in Mathematics}}
  (\bibinfo{publisher}{American Mathematical Society},
  \bibinfo{address}{Providence, RI}, \bibinfo{year}{1997}).

\bibitem[{\citenamefont{Newman and Girvan}(2004)}]{Newman2004}
\bibinfo{author}{\bibfnamefont{M.~E.~J.} \bibnamefont{Newman}}
  \bibnamefont{and} \bibinfo{author}{\bibfnamefont{M.}~\bibnamefont{Girvan}},
  \bibinfo{journal}{Phys. Rev. E} \textbf{\bibinfo{volume}{69}},
  \bibinfo{pages}{026113} (\bibinfo{year}{2004}).

\bibitem[{\citenamefont{Kuramoto}(1984)}]{Kuramoto1984}
\bibinfo{author}{\bibfnamefont{Y.}~\bibnamefont{Kuramoto}},
  \bibinfo{journal}{Prog. Theor. Phys.} \textbf{\bibinfo{volume}{79}},
  \bibinfo{pages}{223} (\bibinfo{year}{1984}).

\end{thebibliography}

\end{document}